\documentclass[prb,twocolumn,aps]{revtex4-1}
\usepackage{verbatim} 
\usepackage{graphicx}
\usepackage{epstopdf}
\usepackage{epsfig}
\usepackage{amsmath}
\usepackage{amssymb}
\usepackage{wrapfig}

\begin{document}

\title{Photoluminescence and Raman investigation of stability of InSe and GaSe thin films}

\author{O. Del Pozo-Zamudio$^1$}
\email{o.delpozo@sheffield.ac.uk}
\author{S. Schwarz$^1$}
\author{J. Klein$^2$}
\author{R. C. Schofield$^1$}
\author{E. A. Chekhovich$^1$}
\author{O. Ceylan$^2$}
\author{E. Margapoti$^2$}
\author{A. I. Dmitriev$^3$}
\author{G. V. Lashkarev$^3$}
\author{D. N. Borisenko$^4$}
\author{N. N. Kolesnikov$^4$}
\author{J. J. Finley$^2$}
\author{A. I. Tartakovskii$^1$}
\email{a.tartakovskii@sheffield.ac.uk}
\affiliation{$^1$Department of Physics and Astronomy, University of Sheffield, Sheffield S3 7RH, United Kingdom}
\affiliation{$^2$Walter Schottky Institut and Physik Department, Technische Universit\"{a}t M\"{u}nchen, Am Coulombwall 4, 85748 Garching, Germany}
\affiliation{$^3$I. M. Frantsevich Institute for Problems of Material Science, NASU, Kiev-142, Ukraine}
\affiliation{$^4$Institute of Solid State Physics, RAS, Chernogolovka 142432 Russia}

\date{\today}

\begin{abstract}

Layered III-chalcogenide compounds belong to a variety of layered crystals that can be implemented in van der Waals heterostructures. Here we report an optical study of the stability of two of these compounds: indium selenide (InSe) and gallium selenide (GaSe). Micro-photoluminescence (PL) and Raman spectroscopy are used to determine how the properties of thin films of these materials change when they are exposed to air at room temperature. We find that in GaSe films, PL signal decreases on average below 50\% over 24 (72) hours of exposure for films with thicknesses 10-25 (48-75) nm. In contrast, weak PL decrease of less than 20\% is observed for InSe nm films after exposure of 100 hours. Similar trends are observed in Raman spectroscopy: within a week, the Raman signal decreases by a factor of 10 for a 24 nm thick GaSe, whereas no decrease was found for a 16 nm InSe film. We estimate that when exposed to air, the layers adjacent to the GaSe film surface degrade and become non-luminescent with a rate of 0.14$\pm$0.05 nm/hour. We show that the life-time of the GaSe films can be increased by up to two orders of magnitude (to several months) by encapsulation in dielectric materials such as SiO$_2$ or Si$_x$N$_y$.

\end{abstract}

\maketitle

\section{Introduction}

Since the isolation of graphene monolayers \cite{novoselov2004}, a large variety of layered compounds has emerged \cite{NovoselovPNAS,XuChemRev}. Among others, some metal-chalcogenides (MCs) show semiconducting properties with sizable band-gaps\cite{Wang}. Recent progress in fabrication of so-called van der Waals heterostructures opens up new possibilities for two-dimensional (2D) MC films in nano- and opto-electronics \cite{geim2013van,Withers}. Layered III-chalcogenides are van der Waals crystals having direct-band gaps in the visible (e.g. GaS, GaSe, GaTe) and near infrared (e.g. InSe, InTe, InS). Thanks to their optical properties and controllable n- or p-type doping\cite{Mudd3}, these materials possess significant promise for van der Waals heterostructures, and will enable a range of band alignments and potential profiles when combined with graphene and other 2D crystals. In contrast to transition metal dichalcogenides (TMDCs) where light emission occurs only in films with a single unit cell thickness \cite{Wang,Splendiani,Mak}, III-VI materials are bright light emitters in a range of thicknesses \cite{Mudd,DelPozo} and offer useful properties for applications\cite{Liu,Wang,Tang,Late} in the emerging 2D materials-based technology. 

An important issue for a variety of layered materials is stability of thin films in ambient conditions\citep{Island,Favron,Bianco,Butler,XuChemRev}. This has particularly showed significance in recent studies of semiconducting black phosphorus \cite{Island}, MoTe$_2$\cite{Chen} and layered superconductors such as NbSe$_2$\cite{ElBana}. We show in this work that film stability issues have to be taken into account for III-VI films, in particular GaSe. 

Both InSe and GaSe are layered crystals with strong covalent in-plane inter-atomic bonding and weaker van der Waals inter-plane bonding \cite{Camara,Capozzi}. The single tetralayer having hexagonal in-plane structure consists of two Se atoms and two In or Ga atoms: Se-Ga-Ga-Se and Se-In-In-Se as shown in Fig.\ref{fig:InSeimage}(a) where the crystal structure of both materials is presented. InSe is a direct gap semiconducting material in its bulk form and has important applications in photo-voltaic devices \cite{Afzaal}. More recently it has been studied in its layered form \cite{Mudd,Mudd2,Sanchez}, where tuning of its band-gap by varying the film thickness has been demonstrated making this material promising for van der Waals heterostructures such as light emitting diodes \cite{Withers}. GaSe is well known for its nonlinear properties\cite{Auerhammer} and in the past few years thin films of this material were used for fabrication of photo-detectors \cite{Hu} and in optical microcavities, where control of light-matter interaction was demonstrated for two-dimensional films\cite{Schwarz}.


In this work we investigate the stability of mechanically exfoliated thin films of InSe and GaSe with thicknesses ranging from 9 to 75 nm using micro-photoluminescence ($\mu$PL) and micro-Raman spectroscopy. We observe decrease of both the PL and Raman signal over a few days. Between the optical measurements the films are kept in air at room temperature. We interpret the decrease of both signals as gradual erosion of the crystal due to interaction with oxygen and water in the atmosphere. This results in significant changes of the crystal properties, leading to modification and decrease of the Raman signal, and also weaker PL following occurrence of non-radiative centers at the layers adjacent to the films surfaces \cite{DelPozo}. For GaSe, we estimate the rate of erosion of 0.14$\pm$0.05 nm/hour based on our previous detailed measurements of thickness-dependent PL, where we observed a strong reduction of PL intensity with the decreasing film thickness\cite{DelPozo}. Here we also demonstrate that encapsulation of the films in SiO$_2$ or Si$_x$N$_y$ allows to overcome the problem of film degradation and prolongs the life-time of the films by two orders of magnitude up to several months. 

\section{Samples and experimental methods}

Bulk GaSe studied in this work was grown by high-pressure vertical zone melting in graphite crucibles under Ar pressure from high-purity materials:  Ga - 99.9999 \% and Se - 99.9995 \%  \cite{Kolesnikov1,Borisenko1}. InSe single crystals were grown by the Bridgman-Stockbarger method from a preliminarily synthesized ingot \cite{Dmitriev}. The conductivity of both materials is n-type, a typical observation in selenides.

\begin{figure}
\begin{center}
\includegraphics[width=0.9\linewidth]{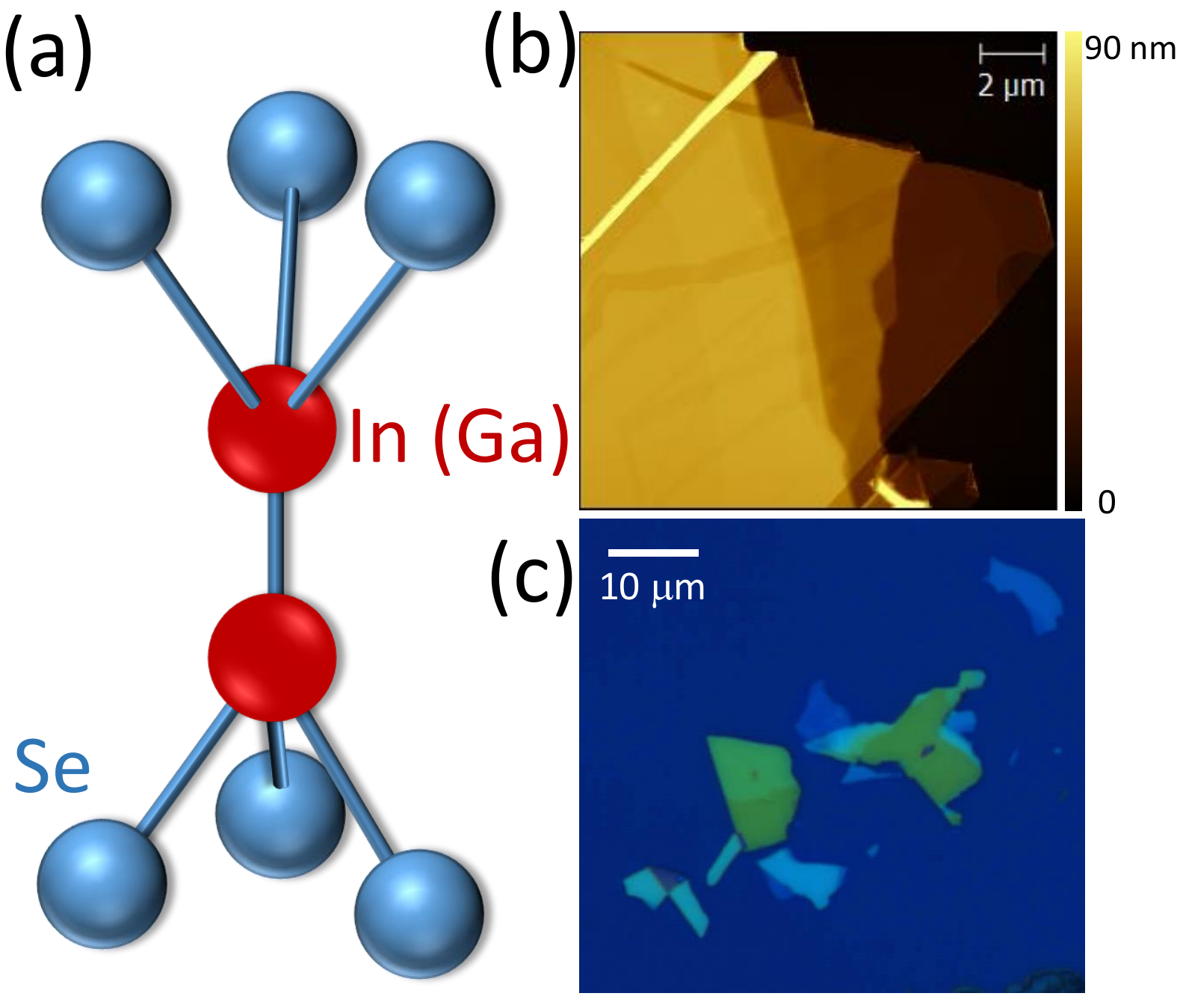}
\caption{(a) Crystal structure of InSe and GaSe. (b) AFM image of a multi-layer GaSe film. (c) Optical image of various isolated InSe thin films on SiO$_2$/Si substrates. Different colors in the image are associated with different film thicknesses.}
\label{fig:InSeimage}
\end{center}
\end{figure}

InSe and GaSe films were exfoliated from bulk using a mechanical cleavage method and were deposited on thermally oxidized silicon substrates. This process generates a variety of films with different thicknesses. Encapsulated GaSe samples were fabricated with the same method and completed with additional deposition  of a 10 nm capping layer of Si$_x$N$_y$ or SiO$_2$ by plasma-enhanced chemical vapor deposition (PECVD). Atomic force microscopy (AFM) was used to measure the thickness of all studied films, including the capped structures. In Fig. \ref{fig:InSeimage}(b) an AFM image of a multi-layer GaSe film is presented showing that the material thickness varies from $\sim$20 nm to $\sim$80 nm across the film (see the scale on the right). In \ref{fig:InSeimage}(c) a group of InSe films with various thickness is presented in an optical microscope image. Here, film areas with different thicknesses are observed as different colors\cite{CastellanosAPL}. As evidenced from both images, by using a laser spot with a diameter of $\approx 1.5-2\mu$m, film areas with constant thickness can be reliably addressed in micro-PL and micro-Raman experiments. 

Optical characterization of GaSe and InSe thin films was carried out using two different techniques: micro-PL ($\mu$PL) and micro-Raman spectroscopy. The $\mu$PL set-up was equipped with a 532 nm continuous wave (cw) semiconductor diode laser for excitation of the sample placed in a flow cryostat at $T$=10K. The laser spot on the sample was $\approx 2 \mu$m. PL signal was collected with an NA=0.55 objective and detected with a 0.5 m single spectrometer and a liquid nitrogen cooled charge coupled device. Micro-Raman was measured at room temperature using an InVia Renishaw System using a 532 nm laser for excitation. 

We study PL signal in GaSe and InSe films following their exposure to air starting from 1 hour after sample fabrication up to 100 hours. The time the sample is kept in the cryostat during the cool-down, the experiment itself and warm-up is not counted, as during this period the sample is kept in vacuum. After the measurement is complete and the sample returns to room temperature in the cryostat, it is taken out and stored in ambient conditions. The measurements in the cryostat were repeated after keeping the sample in air for 24 hours with a maximum total exposure of 100 hours. Raman experiments were carried out on the day of fabrication within a few hours after exfoliation and one week later.

\section{Results}

\begin{figure}
\begin{center}
\includegraphics[width=\linewidth]{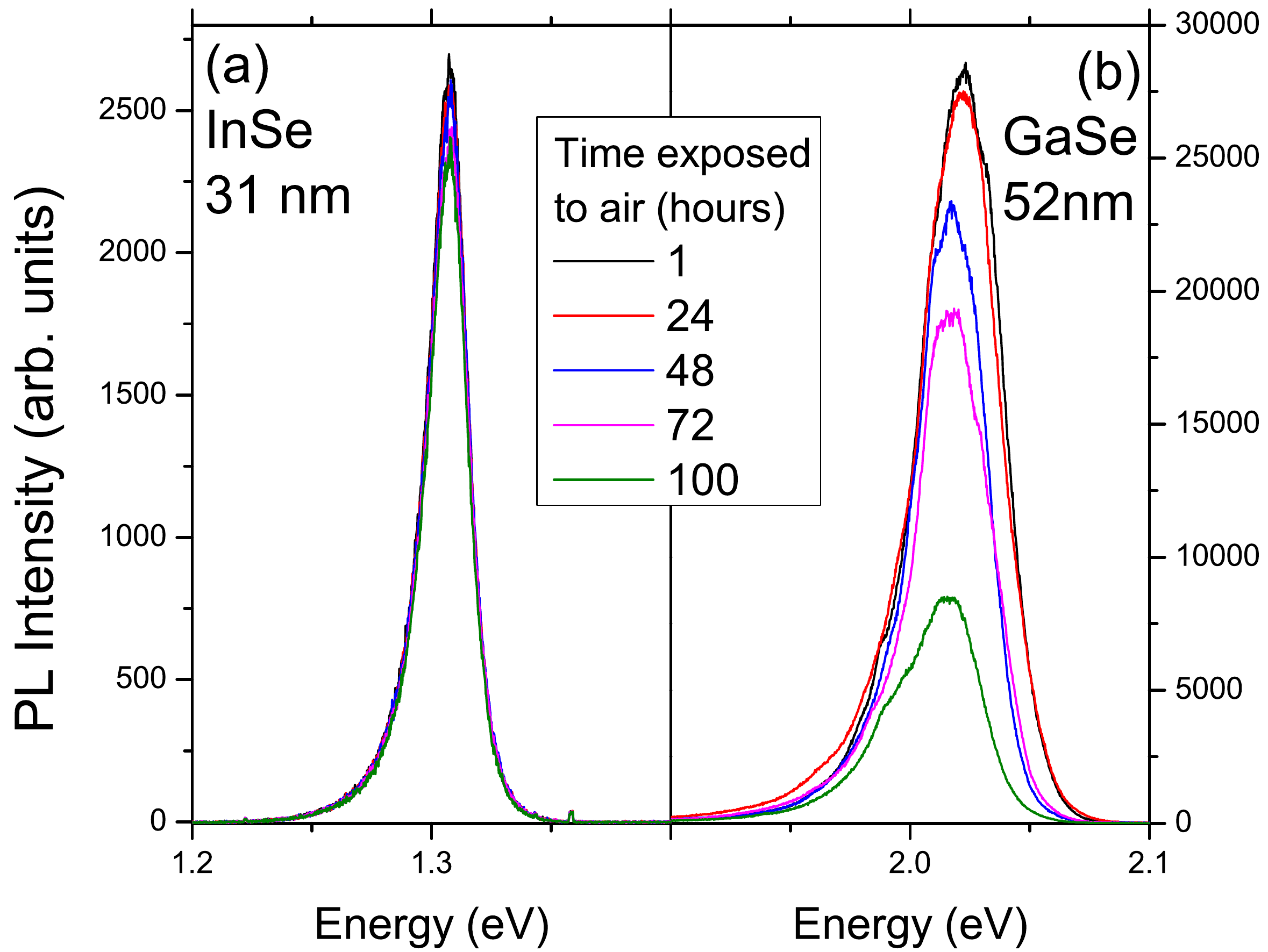}
\caption{Micro-photoluminescence spectra measured at low temperature of 10 K for (a) a 31 nm InSe film and (b) a 52 nm GaSe film after exposure to air from 1 to 100 hours.}
\label{fig:PLspec}
\end{center}
\end{figure}

Fig.\ref{fig:PLspec} shows low-T $\mu$PL spectra of (a) a 31 nm InSe and (b) a 52 nm GaSe films measured with 1 mW laser power. The spectra are recorded from the same area of both films after different exposure to air. The InSe film shows a typical excitonic peak at $\sim$1.3 eV in agreement with previous reports \cite{Mudd2,Sanchez,Zolyomi}. PL intensity exhibits a negligible decrease from measurement to measurement that can be attributed to the re-alignment accuracy. For GaSe, a peak at $\sim$2.02 eV is observed that we relate to impurity/defect states, which typically show bright PL in this material \cite{DelPozo}. In the material studied in this work, the free exciton feature is only observed under pulsed excitation when the impurity states are saturated \cite{DelPozo}. In contrast to the InSe film in Fig.\ref{fig:PLspec}(a), for GaSe, PL signal changes its spectral shape and decreases significantly after the film has been exposed to air for several days. 

\begin{figure}
\begin{center}
\includegraphics[width=\linewidth]{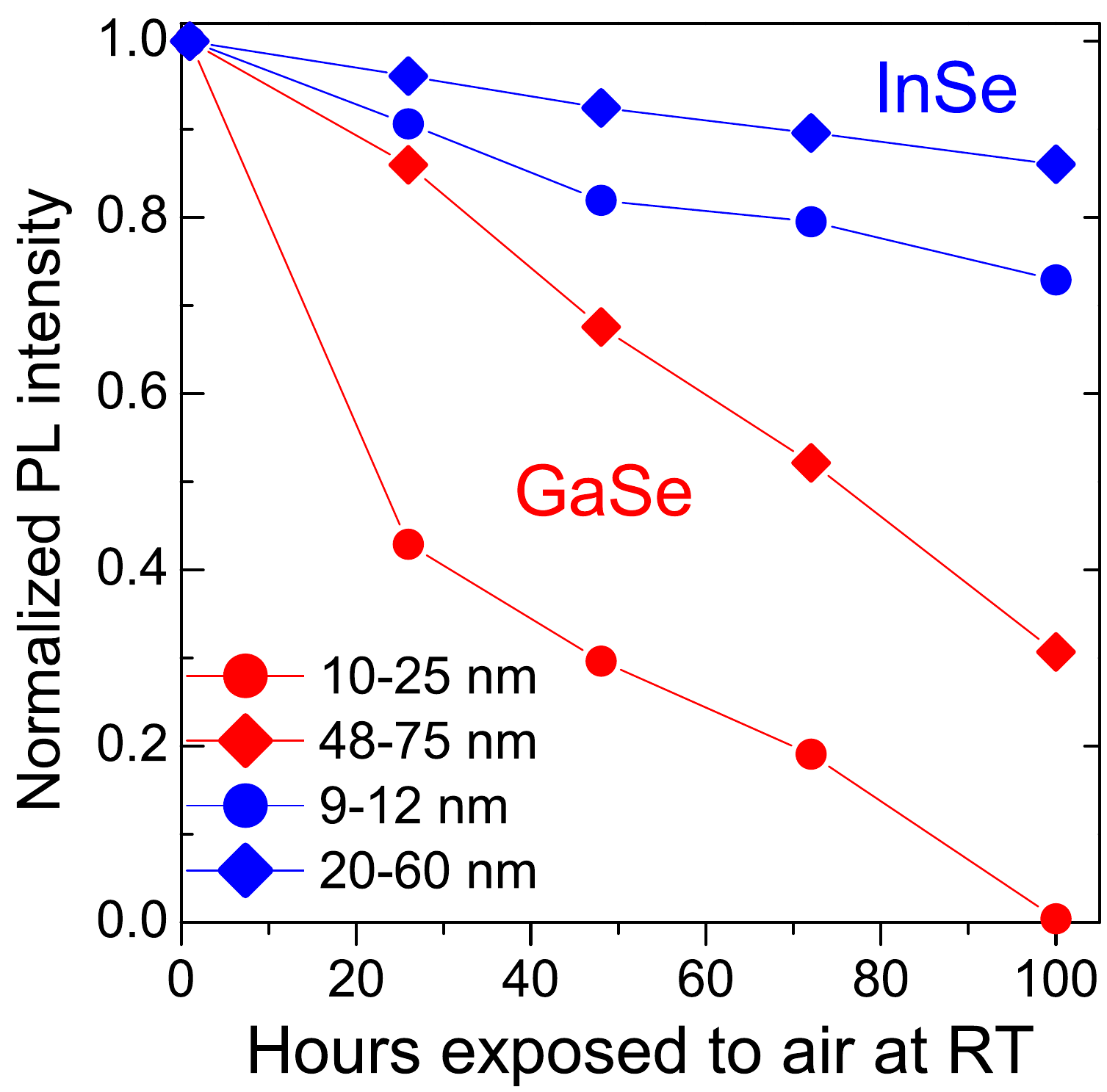}
\caption{(a) PL intensity measured at $T$=10 K as a function of time $t_{exp}$ the InSe and GaSe thin films were exposed to air at room temperature. For InSe, the results are averaged and normalized by the intensity for $t_{exp}$=1 hour for two groups of five films with thicknesses in the range 9-12 nm (blue dots) and 20-60 nm (blue squares). For GaSe the same procedure is carried out for films with thicknesses in the range of 10-25 nm (red dots) and 48-75 nm (red squares).}
\label{fig:PLstab}
\end{center}
\end{figure}

Similar PL experiment was performed for 10 thin films of different thicknesses for each material. Figure \ref{fig:PLstab} shows a comparison of integrated PL intensity of these InSe and GaSe thin films normalized by their intensity during the first measurement (1 hour after exfoliation). In order to reveal typical trends, we averaged over two groups of films showing similar behavior: for group 1 we selected relatively thick films, 20-60 nm for InSe and 48-75 nm for GaSe; for group 2 we included thin films of 9-12 nm for InSe and 10-25 nm for GaSe. For group 1 InSe films (blue squares) the PL intensity decreases by only about 12 \% after 100 hours of exposure to air. For group 2 InSe films (blue dots) during the same period of time it decreases by a factor of two more, 25 \%. For the GaSe films (red squares), we find that after just 24 hours exposed to air the intensity reduction is 15 \% for group 1 thick films and 60 \% for the thin films (red dots). Such rapid reduction of PL intensity is observed for both types of GaSe films, with the strongest effect for the case of the thin films where PL completely vanishes after 100 hours of exposure to air, whereas for the thick GaSe films, PL falls to 30 \% of its initial intensity.

\begin{figure}
\begin{center}
\includegraphics[width=0.8\linewidth]{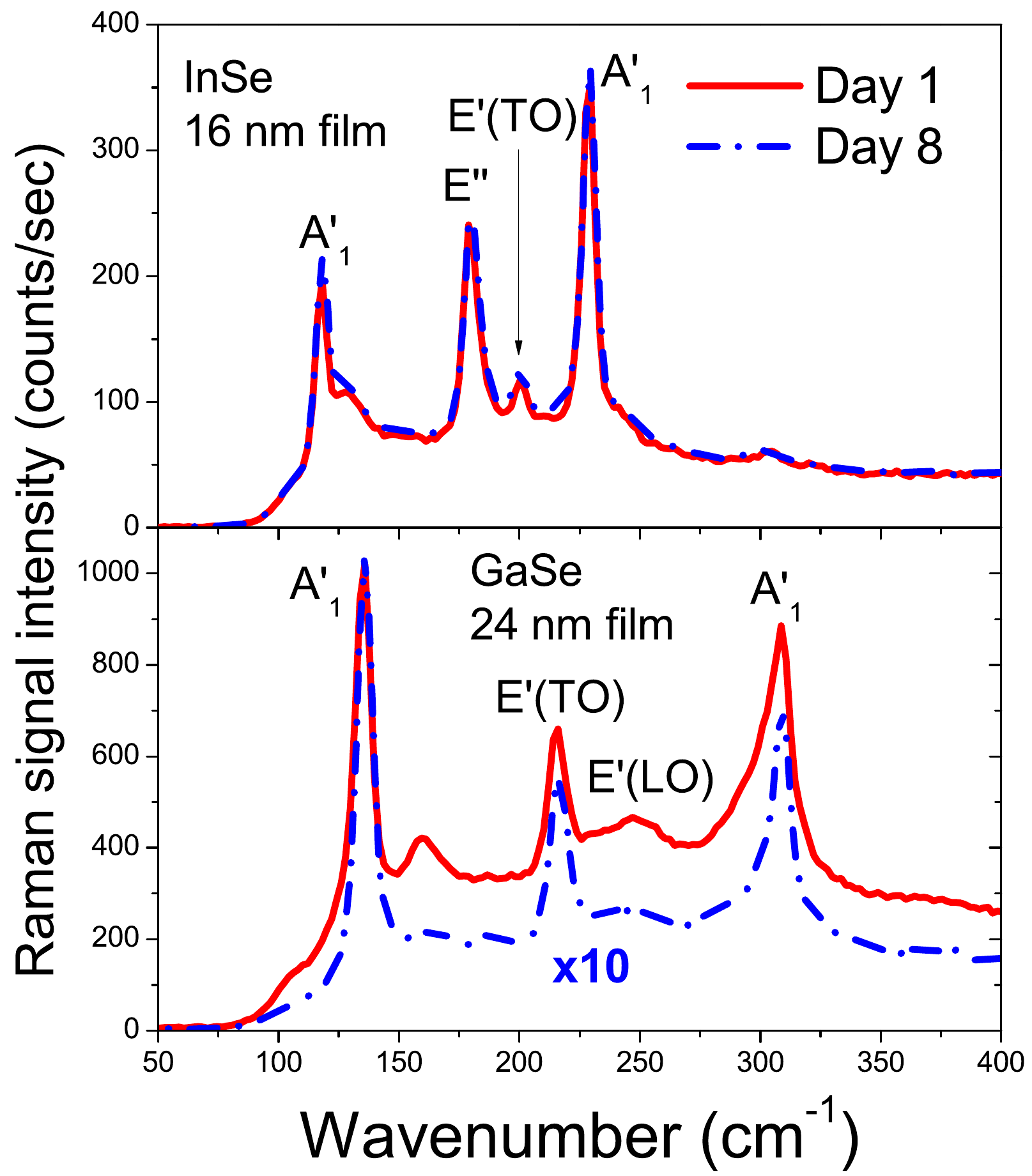}
\caption{Room temperature Raman spectra for a 16 nm InSe (top) and a 24 nm GaSe (bottom) film. The spectra were measured right after fabrication (solid line) and one week later (dashed line).}
\label{fig:Raman}
\end{center}
\end{figure}

Similar behaviour of the signal intensities from the two materials is observed in Raman spectroscopy. In Fig. \ref{fig:Raman} Raman spectra of a 16 nm InSe and a 24 nm GaSe films are presented measured at room temperature. Typical modes reported previously for bulk GaSe and InSe \cite{AllakhverdievRaman,Rodriguez,Kuroda,Kuroda2,Sanchez} are observed for these thin films. For InSe (top panel), the prominent peaks at 117.5, 179.5, 201.7 and 228.6 cm$^{-1}$ correspond to the the vibrational modes A'$_1$, E'', E'(TO) and A'$_1$ respectively. In the case of GaSe (bottom panel) the modes A'$_1$, E'(TO), E'(LO) and A'$_1$ at 135.6, 215.6, 308.7 cm$^{-1}$ respectively, are observed. These Raman modes were clearly observed in both samples on the day when the samples were fabricated. The measurements were then repeated one week later, during which time both samples were stored at room temperature in air. As seen in the figure, the Raman intensity for InSe remains unchanged. In contrast, the Raman spectra for GaSe show significant variation with time: (i) the intensity drops by a factor of 10 and (ii) different relative strengths of the Raman modes are observed. These changes indicate significant modification of physical properties of GaSe films after exposure to air for 1 week.

\section{Discussion}

The significant decrease of PL and Raman signals for GaSe can be attributed to the interaction of the film layers with water and oxygen in the atmosphere, similar to recent reports on black phosphorus \cite{Favron,Island}. Such processes occur in the layers adjacent to the surface(s) of the film, leading to a significant change of the material properties. In a simplified way, this can be described as reduction of the effective film thickness with time when it is exposed to air, as it can be reasonably assumed that only the unaffected material would give rise to typical PL and Raman signals, whereas the eroded GaSe would show very low PL intensity (no PL emission at new wavelengths compared with the original samples were found). 

Using our previous results for the detailed thickness-dependence of PL intensity in GaSe thin films \citep{DelPozo}, we can estimate the time-dependent changes of the effective film thickness in our present experiments. First, the PL dependence on exposure time for each film was analyzed and effective thicknesses were obtained for each exposure time. This was carried out by matching the observed changes in PL intensities with those previously measured for a large number of films of different thicknesses, where we observed a strong reduction of PL intensity with the decreasing film thickness\cite{DelPozo}. By performing this procedure, we deduced the erosion rate for each of the ten films, and finally obtained an average rate for all films of 0.14$\pm$0.05 nm/hour. 

In contrast, for InSe we suggest that the crystal erosion occurs just for a small number of atomic layers at the surface and then possibly saturates causing insignificant impact on PL and Raman intensities. 

The fact that GaSe thin films are sensitive to the environmental effects highlights that additional techniques are required in order to protect this material from interaction with atmosphere. Alternative methods such as fabrication of samples in an inert atmosphere as well as encapsulation of the films with dielectric materials like boron nitride (hBN)  \cite{Cao}, SiO$_2$ or Si$_x$N$_y$\cite{Sercombe,Kretinin} is required. For example, thin layers of black phosphorus exposed to air have been reported to degrade over a few minutes under environmental influences such as humidity, oxygen and exposure to visible light \cite{Island,Castellanos}. However, such processes have been reported to be suppressed if films are exfoliated in an oxygen-free atmosphere \citep{Cao, Favron}. 

\begin{figure}
\begin{center}
\includegraphics[width=0.8\linewidth]{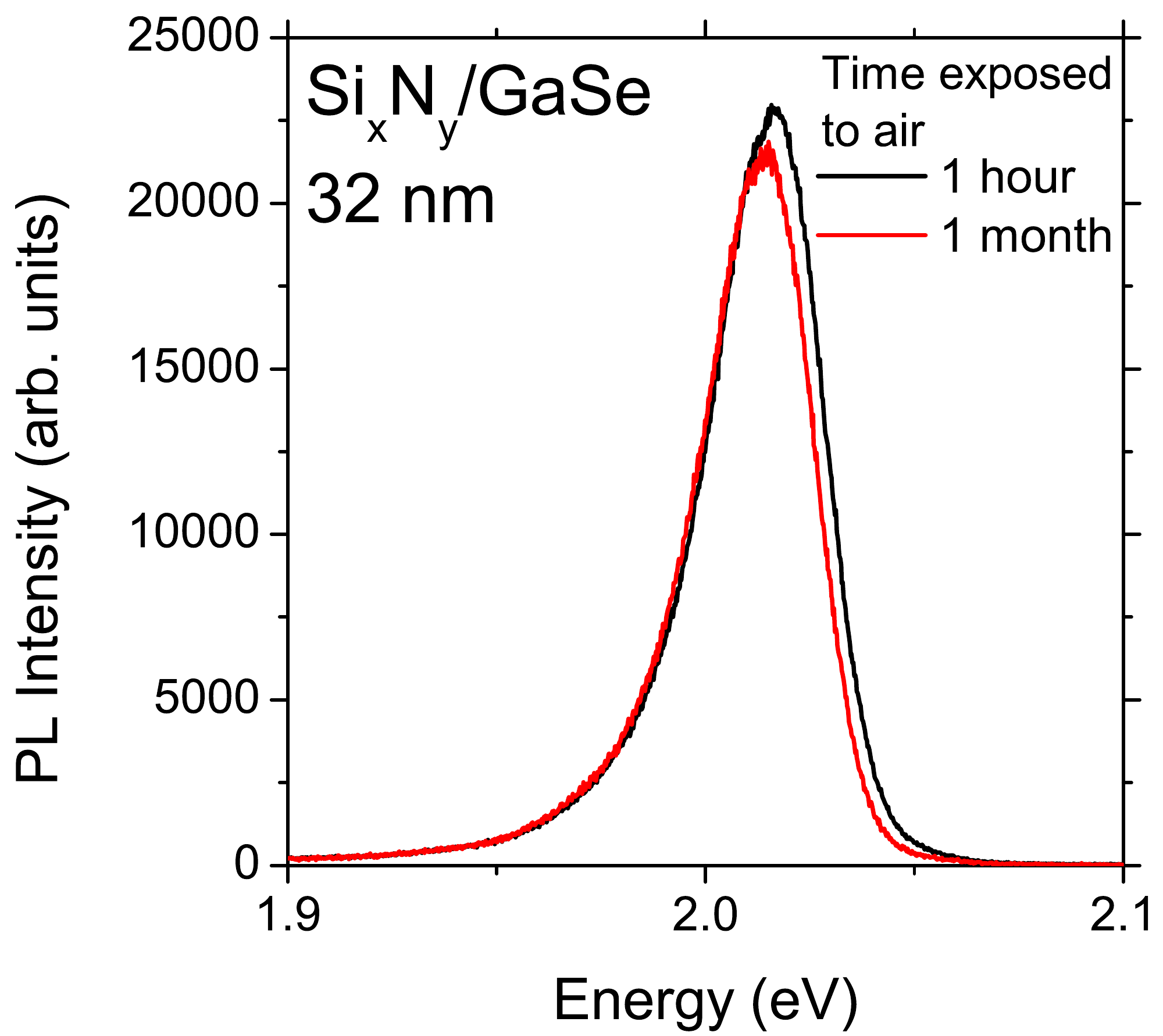}
\caption{PL spectra measured at $T$=10 K for a 32 nm thick GaSe film capped by a 10 nm PECVD layer of Si$_x$N$_y$. The film was measured one hour (black) and one month (red) after fabrication showing negligible signal decay.}
\label{fig:GaSecap}
\end{center}
\end{figure}

In this work, we fabricated a large number of GaSe films of different thicknesses capped with 10 nm Si$_x$N$_y$ or SiO$_2$. The films were initially produced by mechanical exfoliation in air, after which PECVD deposition was carried out. In Fig.\ref{fig:GaSecap} $\mu$PL spectra of a 32 nm GaSe film capped with Si$_x$N$_y$ are presented. The spectrum shown in black was acquired one hour after fabrication, and the spectrum shown with a red curve was measured one month later. The difference between the intensity and lineshape of the two spectra is negligible, indicating significant slowing of the degradation processes. Similar degradation slow down was observed for films of any thickness for either Si$_x$N$_y$ and SiO$_2$ capping.

\section{Conclusions}

Our results show that chemical interactions in the normal atmosphere lead to erosion of GaSe layers adjacent to the film surface with an average rate of 0.14$\pm$0.05 nm/hour, leading to significant decay over time of PL and Raman intensity. Such degradation is significantly weaker for InSe thin films. We relate this behaviour of GaSe with surface oxidation and interaction with water in the atmosphere. Our findings imply that InSe will be the least demanding III-VI layered material for the use in devices based on van der Waals heterostructures, as it requires minimum protection from atmosphere. GaSe thin films are in contrast relatively unstable under ambient conditions. However, the stability of these films can be significantly improved using encapsulation with dielectrics such as Si$_x$N$_y$ or SiO$_2$, which slow down the degradation rate by more than two orders of magnitude.

\section*{Acknowledgments}

We thank the financial support of EPSRC grant EP/M012700/1, Graphene Flagship project 604391, FP7 ITN S3NANO, EPSRC Programme Grant EP/J007544/1 and CONACYT.  We are grateful to Fa. Soliton for the access to Raman facilities.\\ 

\bibliography{references}

\end{document}